\documentclass[aps,eqsecnum,floats,twocolumn,amssymb,amsmath,showpacs]{revtex4}

\usepackage{graphicx}
\begin{document}
\title{Soft elasticity of RNA gels and negative Poisson ratio}

\author{Amir Ahsan, Joseph Rudnick and Robijn Bruinsma}

\affiliation{Department of Physics, UCLA, Box 951547, Los Angeles,
CA 90095-1547}

\date{\today}

\begin{abstract}

We propose a model for the elastic properties of RNA gels. The model predicts anomalous elastic properties in the form of a negative Poisson ratio and shape instabilities. The anomalous elasticity is generated by the non-Gaussian force-deformation relation of single-stranded RNA. The effect is greatly magnified by broken rotational symmetry produced by double-stranded sequences and the concomitant soft modes of uniaxial elastomers.

\end{abstract}

\pacs{62.20.Dc, 87.14.Gg, 82.35.Lr} \maketitle

\section{Introduction} \label{sec:intro}

        The study of the viscoelastic properties of networks of flexible, synthetic polymers has for many years been a central topic of polymer science. The classical Flory theory for the elasticity of rubber and of gels treats these systems as networks of nodes linked by highly flexible chains\cite{flory,flory1,rubinstein}. Scaling relations for the viscoelastic moduli that result from this model have been well confirmed\cite{rubinstein}. The study of networks of biopolymers has provided a fresh impetus to the field. Gels of semi-flexible biopolymers, like F-Actin, were shown to obey novel scaling relations\cite{gardel}. The focus of the present paper is on the elasticity of a  different biopolymer system, namely a network or gel of RNA chains. The folding of smaller RNA molecules has been already extensively discussed in the molecular biology literature in the context of Ribozymes\cite{thirumalai}, but extended RNA gels have not received much attention. The genome of single-stranded (ss) RNA viruses for instance may form a very promising small scale realization of an RNA gel, as discussed in the conclusion. It is the purpose of this paper to present a simple model for RNA gels that indicates that such gels should have rather unusual elastic properties that will distinguish them not only from Flory-type gels but also from gels of semi-flexible biopolymers like F-Actin. 
        
\begin{figure}[htbp]
\begin{center}
\includegraphics[width=2in]{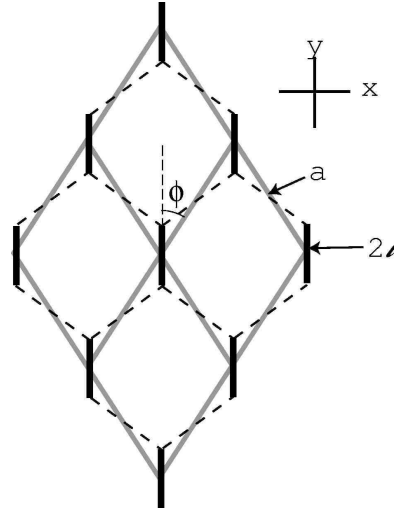}
\caption{The network of rods and springs that forms the basis of the negative Poisson ratio system. As indicated in the figure, the length of the sides of the underlying altered cubic lattice is $a$ and the length of the rods is $2l$. }
\label{fig:rodsandsprings}
\end{center}
\end{figure}

	A single-stranded (``ss'') RNA chain can be folded first into a ``secondary'' structure that consists of the pattern of optimal pairing of the bases of the chain\cite{thirumalai}. This secondary structure is represented as a planar, branched graph of duplexed double-stranded (``ds'') sequences linked by ``bubbles'' and ``stem-loops'' composed of unpaired bases. A three-dimensional ``tertiary'' structure is obtained if one also allows complementary pairing between the bases of different bubbles and stem-loops of the secondary structure. The model we will study assumes a highly simplified tertiary network topology composed of an array of rigid rods (the ds complementary sequences) that are linked by flexible chains (the ss sequences). Two flexible chains emerge from either of the two ends of each rod (see Fig. \ref{fig:rodsandsprings}). Models of this type have in fact been used to describe the folding kinetics of Ribozymes \cite{isambert}. A key ingredient of our model is that these flexible chains are \emph{not} assumed to have the elastic properties of either a Gaussian chain or a semi-flexible worm-like chain. Instead, we will examine the elasticity of the network for a general class of interaction potentials. Specifically, we will consider chains that obey the force-extension curve of ss DNA as measured by single-molecule micromechanics.

The particular network of rods and springs that we propose to investigate as a representation of complexed RNA  is displayed in Fig. \ref{fig:rodsandsprings}.
The rods are centered on an altered square lattice. The angle, $\phi$ that an edge of this lattice makes with respect to the vertical, as shown in Fig. \ref{fig:rodsandsprings}, parameterizes the overall characteristics of the lattice. If $\phi$, which ranges between 0 and $\pi/2$, is equal to $\pi/4$ the lattice is square. At either of the limits of $\phi$, the lattice has collapsed onto itself. The edges of the unit cells of the lattice, which do not represent any physical quantity, are represented by grey lines in the figure. The rods are thick solid lines, and the springs are depicted as dashed lines connecting the ends of neighboring rods. As shown in the figure, the length of the edges of the underlying lattice is $a$, and the length of the rods is $2l$. 

The model belongs to a class of systems, namely uniaxial and biaxial elastomers, that are known to have unusual elastic properties. The internal rotational degrees of freedom produce what is known as ``soft'' elasticity in the form of large shape changes under applied fields as well as a vanishing of the Poisson ratio \cite{warnerkutter}. The anomalous elasticity of networks with broken rotational symmetry is an unavoidable and fundamental feature according to a theorem by Golubovic and Lubensky \cite{golublubensk}. One of the aims of the present paper is to examine the consequences of this theorem for a concrete model. Specifically, we will explore precisely what physical properties of the interaction potential characterizes anomalous soft elasticity. We will argue in particular that control of the soft elasticity can be achieved by altering the physical properties of the non-Gaussian springs and that this can be achieved in the context of RNA gels.  

\begin{figure}[htbp]
\begin{center}
\includegraphics[width=2in]{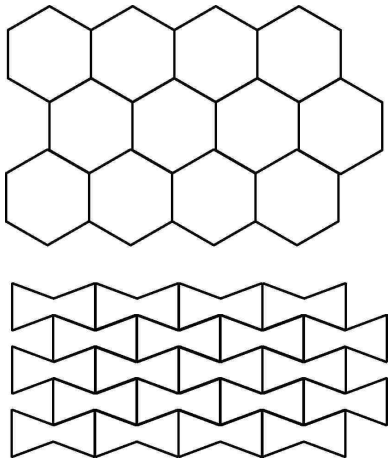}
\caption{Two types of hexagonal networks of hinged rigid rods. The network shown at the top of the figure exhibits a positive Poisson ratio, in that it expands vertically when compressed horizontally. The network at the bottom contracts vertically when compressed horizontally and thus has a negative Poisson ratio.}
\label{fig:networks}
\end{center}
\end{figure}

The proposed model can be viewed as a modified version of a hexagonal lattice of hinged rods. It is actually known in the material science literature that systems that can be represnted by models of this type exhibit unusual elasticity. A conventional hexagonal lattice, shown at the top Fig. \ref{fig:networks}, will expand vertically when compressed horizontally if the rods are rigid and freely hinged. 
On the other hand, the network at the bottom of the figure---consisting of non-convex hexagonal units---contracts vertically when compressed horizontally. The network is thus characterized by a \emph{negative} Poisson ratio. The conditions for constructing a network of hinged rods that exhibits a negative Poisson ratio can be elegantly demonstrated by the popular Hoberman Sphere \cite{hoberman},  which maintains its spherical shape as it expands and collapses\cite{note}. The generic term for a material with a negative Poisson ratio is ``auxetic."  A negative Poisson ratio is known to be exhibited by Iron Pyrites \cite{love}, self-avoiding, fixed-connectivity membranes \cite{bowick}, monocrystalline Zinc \cite{lubarda}, Carbon Nitride \cite{yuejin}, polyethylene foams \cite{brandel}, two dimensional mesh-like systems \cite{gaspar}, structures composed of rotating rigid units \cite{grima} and in network-embedded composites \cite{evans}. Biologically, a negative Poisson ratio has been found to be exhibited by both skin \cite{skin} and bone \cite{bone}.

\section{Poisson ratio of a two dimensional network of rods and springs} \label{sec:2d}

We will start by exploring the properties of the model shown in Fig. \ref{fig:rodsandsprings} as a two-dimensional network. We will represent a distortion of the lattice in terms of a strain tensor $\stackrel{\leftrightarrow}{\epsilon}$, so that the displacement $\Delta \vec{r}_i = ( \Delta x_i, \Delta y_i )$, of a lattice vertex originally at the location $\vec{r}_i=(x_i, y_i)$ is given by
\begin{equation}
\Delta x_i  = \epsilon_{xx} x_i + \epsilon_{xy} y_i 
\label{eq:deltax2d}
\end{equation}
and similarly for $\Delta y_i$. We will also allow the rods to rotate in the plane by the angle $\theta$. 

\begin{figure}[htbp]
\begin{center}
\includegraphics[width=1in]{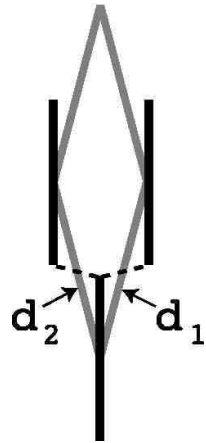}
\caption{A triad of neighboring rods, connected in this figure by two springs connected to the top end of the lower central rod. The rods and the lattice are shown in an as-yet undistorted state. The lattice is compressed horizontally as compared to the one shown in Fig. \ref{fig:rodsandsprings}.}
\label{fig:triad}
\end{center}
\end{figure}

Consider, now, the triad of rods with connecting springs shown in Fig. \ref{fig:triad}.
Note that the underlying lattice is more compressed along the horizontal direction than the lattice shown in Fig. \ref{fig:rodsandsprings}. As will be demonstrated below, the degree of alteration of the square lattice is controlled by the requirement that the rod and spring network can be stabilized by an {\em osmotic} pressure.  The distance between the upper end of the lower rod and the lower end of the rod on the right is equal to
\begin{eqnarray}
d_1^2& = & 
(a \epsilon_{xy} \cos \phi -2 l \sin \theta +a \sin
   \phi +a \epsilon_{xx} \sin \phi )^2 \nonumber \\ &  + &(-2 l \cos
   \theta +a \cos \phi +a \epsilon_{yy} \cos \phi +a
   \epsilon_{yx} \sin \phi )^2 \nonumber \\
   \label{eq:dist1}
\end{eqnarray}
and the distance between the upper end of the lower rod and the lower end of the rod to the left is given by
\begin{eqnarray}
d_2^2 & = &(a \epsilon_{xy} \cos \phi -2 l \sin \theta -a \sin
   \phi -a \epsilon_{xx} \sin \phi )^2 \nonumber \\ & + &(-2 l \cos
   \theta +a \cos \phi +a \epsilon_{yy} \cos \phi -a
   \epsilon_{yx} \sin \phi )^2 \nonumber \\
   \label{eq:dist2}
\end{eqnarray}
The angle $\theta$ in (\ref{eq:dist1}) and (\ref{eq:dist2}) is the angle to which all the rods rotate under the influence of the uniform distortion of the lattice as parameterized by the constant strain tensor $\stackrel{\leftrightarrow}{\epsilon}$.

\subsection{Non-Gaussian behavior of the springs}

In order to facilitate numerical calculations, we will parameterize the interactions mediated by the springs in terms of the square of the distance between the spring end-points. That is, we express the energy of the interaction in terms of the variable $x$, where
\begin{equation}
x = d^2-d_0^2
\label{eq:distance}
\end{equation}
The quantity $d$ is the distance between end points, and $d_0$ is the distance between end-points in the equilibrium state. Next, let $V(d)$ be the interaction potential that represents the free energy of an ss sequence connecting two points a distance $d$ apart. We will expand $V(d)$ in a Taylor series around the equilibrium state $x=0$:
\begin{eqnarray}
\lefteqn{V\left(\sqrt{d_0^2 + x}\right)} \nonumber \\ & = & V(d_0) + \frac{x}{2d_0} V^{\prime}(d_0) + \frac{x^2}{8 d_0^3} \left[-V^{\prime}(d_0) + d_0 V^{\prime \prime}(d_0) \right] \nonumber \\ && + \cdots \nonumber \\
& = & \nu_0 + \nu_1 x + \frac{\nu_2}{2} x^2 + \cdots
\label{eq:expansion}
\end{eqnarray}
It is important to note that even if the interaction, $V(d)$, is convex upward, so that $V^{\prime \prime}(d)>0$, \emph{the second order coefficient, $\nu_2$, in the expansion above can be negative}. Consider for instance the case of a simple power-law interaction energy of the form
\begin{equation}
V(d) = Cd^p
\label{eq:interaction1}
\end{equation}
with $C$ a positive coefficient and $p$ a power greater than one. Then, 
\begin{equation}
V^{\prime}(d_0) = Cpd_0^{p-1}
\label{eq:interaction2}
\end{equation}
and
\begin{equation}
d_0V^{\prime \prime}(d_0) - V^{\prime}(d_0) = Cp(p-2) d_0^{p-1}
\label{eq:interaction3}
\end{equation}
If $p<2$, then, assuming that the coefficient $C$ is greater than zero, the second order coefficient in (\ref{eq:expansion}), $\nu_2$, will be negative. In the case of Gaussian chains with  Hooke-law-type harmonic interaction, $\nu_2$ is equal to \emph{zero}, as are all higher order coefficients in the expansion of the interaction in terms of the variable $x$. On the other hand, for a freely jointed chain or a worm-like chain $\nu_2 >0$

Under the assumption that the interaction mediated by the springs is always attractive, the angle, $\theta$ at which the rods tilt must adjust in such a way as to minimize the sum of the two distances, $d_1$ and $d_2$. Taking the derivative of $d_1+d_2$ with respect to $\theta$ and constructing an extremum equation, we end up with the relationship
\begin{equation}
8 a l \cos \phi  \left[(\epsilon_{yy}+1) \sin \theta-\epsilon_{xy} \cos \theta \right] =0
\label{eq:thetaeq}
\end{equation}
The strain tensor is assumed to be a small quantity. This means that the solution to the equation above is, to an accuracy sufficient for our purposes
\begin{equation}
\theta = \epsilon_{xy}
\label{eq:thetasol}
\end{equation}

The total energy associated with the interactions mediated by the springs shown in Fig. \ref{fig:triad} is then given by
\begin{eqnarray}
E & = & \nu_0 + \nu_1 \left( \left(d_1^2-d_0^2 \right) + \left(d_2^2-d_0^2 \right) \right) \nonumber \\ &&+ \frac{\nu_2}{2}\left( \left(d_1^2-d_0^2 \right)^2 + \left(d_2^2-d_0^2 \right)^2 \right)
\label{eq:en1}
\end{eqnarray}
with $\theta$ in (\ref{eq:dist1}) and (\ref{eq:dist2}) as given by (\ref{eq:thetasol}). 

\subsection{Determination of the angle $\phi$, expansion of the energy in strain coordinates and calculation of the Poisson ratio}

The next step is to expand the resulting expression to second order in the strain tensor. At zeroth order, we are left with the coefficient $\nu_0$. To first order in $\stackrel{\leftrightarrow}{\epsilon}$ we have the following contribution to the energy
\begin{equation}
4a \nu_1 \left( a \epsilon_{xx}  \sin ^2\phi + \epsilon_{yy}
   \cos \phi  (a \cos \phi -2 l) \right)
   \label{eq:ep1en}
\end{equation}
The appearance of terms linear in the strain energy indicates that the system has a tendency to spontaneously deform. First, consider the case of a uniform contraction. In terms of the strain tensor, this yields an energy going as $\epsilon_{xx} + \epsilon_{yy}$. This tendency will be countered by an osmotic pressure $\Pi$ that increases as the monomer concentration grows under contraction. In order that an osmotic pressure can completely balance the first order energy for $d=d_0$ (the equilibrium state) in (\ref{eq:ep1en}), it is necessary that the coefficients of $\epsilon_{xx}$ and $\epsilon_{yy}$ in that expression be equal to each other. This requirement translates into an equation for the lattice angle $\phi$ that is satisfied when
\begin{equation}
\phi = \arccos \left[ \frac{l}{2a} + \sqrt{\left( \frac{l}{2a}\right)^2 + \frac{1}{2}}\right]
\label{eq:phival}
\end{equation}
If the coefficients of $\epsilon_{xx} $ and $\epsilon_{yy}$ are unequal, then the system undergoes a spontaneous shear tranformation. 
Two limits of the above expression are noteworthy. When $l=0$, so the rods are infinitesimal in extent, then $\phi$ as given by (\ref{eq:phival}) is equal to $\pi/4$, consistent with an underlying square lattice. On the other hand, when $l=a/2$, we find a $\phi$ from (\ref{eq:phival}) that is equal to zero, which implies a horizontal collapse of the complex. In this limit, the rods are exactly long enough that their tips touch in the event of such a collapse of the lattice. Henceforth we will assume that $l$ lies in the range between 0 and $a/2$. 

 Assuming cancellation of the linear term, the energy is now, at lowest non-trivial order, quadratic in the elements of $\stackrel{\leftrightarrow}{\epsilon}$. This contribution to the quadratic energy is of the form
\begin{equation}
\Omega \left[\stackrel{\leftrightarrow}{\epsilon} \right] = \sum_{i,j,k,l} \alpha_{ij,kl}  \ \epsilon_{ij} \epsilon_{kl}
\label{eq:quaden}
\end{equation}
In light of translational invariance, this energy will be the same for the energy supplied by the two springs attached to the upper tip of every rod in the complex, under the assumption of a uniform strain. We have thus effectively calculated the energy of interaction in the entire complex.

The strain induced by an externally applied stress, $\stackrel{\leftrightarrow}{\sigma}$, is the solution to the set of equations
\begin{equation}
\frac{\partial }{\partial \epsilon_{ij} }\Omega \left[\stackrel{\leftrightarrow}{\epsilon} \right]  = \sigma_{ij}
\label{eq:epeqs}
\end{equation}
We will take the strain to be exerted in the $y$ direction, so $\sigma_{yy}$ is the only non-zero element of the strain tensor. After solving (\ref{eq:epeqs}) for the elements of the strain tensor we can extract the Poisson ratio, $\mu$, via the relationship
\begin{equation}
\mu = - \frac{\epsilon_{xx}}{\epsilon_{yy}}
\label{eq:2dpoissdef}
\end{equation}
Following a straightforward calculation we obtain the explicit result
\begin{equation}
\mu = \frac{a^2(a^2-4l^2) \nu_2}{(a^2-l^2 +al \sqrt{2+(l/a)^2})\nu_1   + a^2(a^2-4l^2) \nu_2 }
\label{eq:2dpoissresult}
\end{equation}

\subsection{Poisson ratio when there are no rods}

It is useful to consider certain limiting cases of Eq. (\ref{eq:2dpoissresult}). First, consider the case $l=0$, which corresponds to a network in which the rods are replaced by point contacts. If we set $l=0$ in the expression (\ref{eq:2dpoissresult}), the expression for the Poisson ratio further simplifies to
\begin{equation}
\frac{a^2\nu_2/\nu_1}{1+a^2 \nu_2/\nu_1}
\label{eq:2dpoissl0}
\end{equation}
which predicts a negative Poisson ratio, if the ratio $\nu_2/\nu_1$ is negative. To check for stability of the lattice one can determine the eigenvalues of the energy matrix $\stackrel{\leftrightarrow}{\alpha}$. When $l=0$, this four-by-four matrix is degenerate, and there are two distinct eigenvalues: $a^2 \nu_1 -a^4 \nu_2$ and $a^2 \nu_1 + 3 a^4 \nu_2$. Both eigenvalues are positive, corresponding to a stable lattice, provided $-\nu_1/3a^2 < \nu_2 < \nu_1/a^2$, where we assume that $\nu_1>0$. For $\nu_2$ in this range, the Poisson ratio varies from -1/2 to 1/2. It follows that a moderately negative Poisson ratio is, in fact,  possible in a lattice consisting entirely of non-Gaussian springs. Note that this is not related to the existence of internal rotational degrees of freedom. 

\subsection{Poisson ratio at general $l$}

\begin{figure}[htbp]
\begin{center}
\centerline{\includegraphics[width=2in]{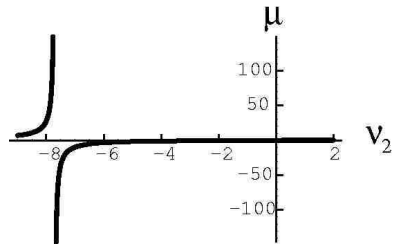}}
\caption{The two Poisson ratio of the two-dimensional lattice according to (\ref{eq:2dpoissresult}). }
\label{fig:poiss2d}
\end{center}
\end{figure}

We now set $l=0.45 a$. This leads to an equilibrium state that is compressed horizontally as  in Fig. \ref{fig:triad}. In Fig. \ref{fig:poiss2d} we show $\mu$ as a function of $\nu_2$.

\begin{figure}[htbp]
\begin{center}
\centerline{\includegraphics[width=2in]{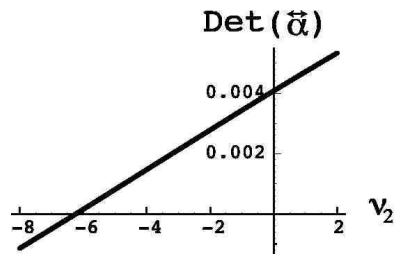}}
\caption{The determinant of the elastic energy matrix $\stackrel{\leftrightarrow}{\alpha}$ when $l=0.45 a$, and $\nu_1$ and $a$ have been set equal to one. }
\label{fig:2ddet}
\end{center}
\end{figure}

\begin{figure}[htbp]
\begin{center}
\includegraphics[width=2in]{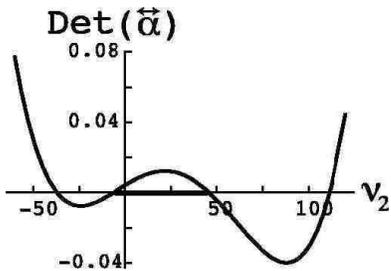}
\caption{The determinant of the energy matrix $\stackrel{\leftrightarrow}{\alpha}$ as a function of $\nu_2$, with $l=0.45 a$ and $a$ and $\nu_1$ set equal to one. The range of stability  is indicated by the heavy line on the horizontal axis.}
\label{fig:extdet}
\end{center}
\end{figure}

We have again checked the stability by computing the determinant of the four-by four matrix with elements $\alpha_{ij,kl}$. Figure \ref{fig:2ddet} shows a typical plot of the determinant as a function of $\nu_2$ .
The determinant vanishes at $\nu_2 = -6.232$, corresponding to the emergence of a negative eigenvalue of the energy matrix $\stackrel{\leftrightarrow}{\alpha}$ and the development of a mechanical instability in the complex \cite{instabnote}. At that threshold value of $\nu_2$, we find $\mu = -4.21226$, so a substantial negative Poisson ratio is indeed possible close to a mechanical instability. Figure \ref{fig:extdet} shows the determinant for an extended range of the interaction coefficient $\nu_2$. As illustrated in that figure, the range of stability of the energy matrix is bounded from above as well as below as  a function of $\nu_2$. 

If we compare these results with those in the case $l=0$ we conclude that the broken rotational symmetry for $l \neq 0$ has greatly { \em amplified} the negative Poisson ratio produced by a negative value of $\nu_2$. There is indeed no intrinsic physical bound on the Poisson ratio in an anisotropic solid \cite{ting}. The amplification effect is most dramatic when $\nu_2$ approaches the threshold of the mechanical instability, at $\nu_2 =-6.232$. Note though that according to Eq. (\ref{eq:2dpoissresult}) the Poisson ratio \emph{formally} is zero even at this critical point for a Gaussian network and any other system with $\nu_2=0$. It should be recalled here that the Poisson ratio is zero in conventional nematic elastomers \cite{warnerkutter}. One can compare the effect of a negative $\nu_2$ to that of a small magnetic field applied to a magnet as one approaches the Curie temperature. The divergence of the susceptibility amplifies the effect of the applied field. In the next section we will extend the notion of the sort of complex we have been discussing to a three dimensional system.

\section{Three dimensional network of rods and springs embodying a negative Poisson ratio} \label{sec:3d}

\subsection{Preliminaries: description of parameters} \label{sec:prelim}

\begin{figure}[htbp]
\begin{center}
\includegraphics[width=2in]{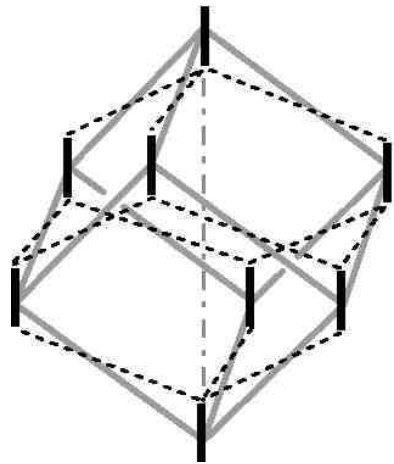}
\caption{Portion of the cubic lattice, shown unsheared here, with the rods at the vertices. The ``springs'' connecting the vertices are shown as dashed lines. The edges of the cubic lattice in which the rods are embedded are shown as grey lines. Note that these edges do not represent actual physical structures. Only those springs that connect two rods shown in the figure are depicted. Finally, a principle diagonal, lying along the $z$ axis, is shown as a dashed and dotted line.}
\label{fig:network3}
\end{center}
\end{figure}

\begin{figure}[htbp]
\begin{center}
\includegraphics[width=3in]{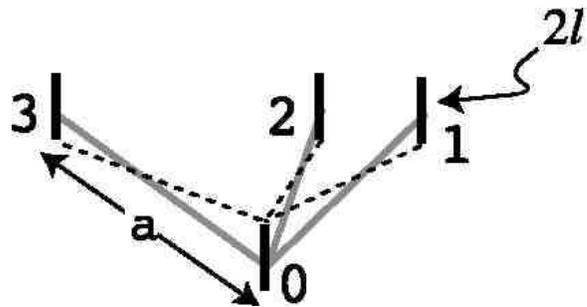}
\caption{Three vertices of the cubic lattice with associated rods and springs attached to them. The numbers are the indices of the vertices. As in Fig \ref{fig:network3}, the springs are shown as dashed lines. }
\label{fig:vertices}
\end{center}
\end{figure}

The way in which the three-dimensional network is constructed will be a generalization of the two-dimensional case. Imagine a sheared version of a cubic lattice rotated so that the $z$-axis is along a diagonal. A rod is placed at each vertex of this lattice, and each end of this rod is connected to the ends of the three rods that are closest to it by a spring having a non-linear force-extension relation. Figure \ref{fig:network3} depicts a portion of this lattice of rods, shown as short solid lines. The springs are represented by dashed lines and the edges of the lattice---shown for illustrative purposes only---are represented by grey lines. 
The $z$ axis is along the principal diagonal, shown as a grey dashed and dotted line in the figure, and the $x$ axis is aligned with the projection in the $x$-$y$ plane of an edge of the cube. Three neighboring vertices of the cube are indexed as shown in Fig. \ref{fig:vertices}.
The organization illustrated there ought to be understood as a kind of ``averaged'' version of a structure in which each rod terminates in two springs, as shown in Fig. \ref{fig:rodsandsprings}.

\begin{figure}[htbp]
\begin{center}
\includegraphics[width=3in]{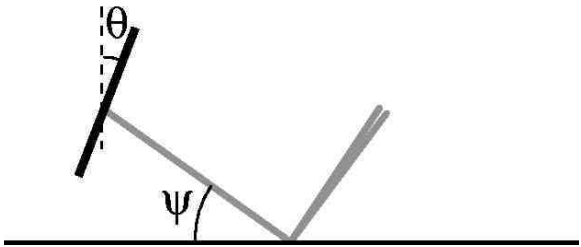}
\caption{The angles $\phi$ and $\theta$. All three edges shown in the figure have an orientation with respect to the horizontal plane. }
\label{fig:angles}
\end{center}
\end{figure}

The rod with index 0 is assumed to be located at the origin. The coordinates of the corners on which the neighboring rods sit are given by 
\begin{eqnarray}
x_1 & = & a \cos \psi \label{eq:x1} \\
y_1 & = & 0 \label{eq:y1} \\
z_1 & = & a \sin \psi \label{eq:z1} \\
x_2 & = & A \cos \psi \cos ( 2 \pi /3) \label{eq:x2} \\
y_2 & = & a \cos \psi \sin ( 2 \pi/3) \label{eq:y2} \\
z_2 & = & a \sin \psi \label{eq:z2} \\
x_3 & = & a \cos \psi \cos( 4 \pi/3) \label{eq:x3} \\
y_3 & = & a \cos \psi \sin( 4 \pi/3) \label{eq:y3} \\
z_3 & = & a \sin \psi \label{eq:z3}
\end{eqnarray}
The quantity $a$ in Eqs. (\ref{eq:x1})--(\ref{eq:z3}) is the length of a cube edge. The quantity $\psi$ is the angle that the edges of the cube in Fig. \ref{fig:vertices} make with respect to the $x$-$y$ plane. In the case of an unsheared cubic lattice, this angle is equal to $\arctan (1/\sqrt{2})$. Note that when $\psi= \pi/2$, the edges connecting the site at the origin to the three neighboring ones are all vertical, so that the neighboring sites all lie at exactly the same location. Each edge in the cubic lattice has length $a$ and the length of the rods is $2l$, again as shown in Fig. \ref{fig:vertices}. Initially, the rods point along the $z$-axis. However, we will assume that when the lattice shears the rods change orientation. The direction in which the rods point will be described in terms of the standard spherical angles $\theta$ and $\phi$. See Fig. \ref{fig:angles} for the representation of the angles $\phi$ and $\theta$. 
The next step is to allow the lattice to distort. As in the previous section we represent the distortion in terms of a strain tensor $\stackrel{\leftrightarrow}{\epsilon}$, so that placing one of the vertices of the lattice at the origin, the displacement $\Delta \vec{r}_i = ( \Delta x_i, \Delta y_i, \Delta z_i)$, of any other lattice vertex originally a the location $\vec{r}_i=(x_i, y_i, z_i)$ is given by
\begin{equation}
\Delta x_i  = \epsilon_{xx} x_i + \epsilon_{xy} y_i + \epsilon_{xz} z_i
\label{eq:deltax}
\end{equation}
and similarly for $\Delta y_i$ and $\Delta z_i$. 

\subsection{Determination of the angles $\psi$, $\phi$ and $\theta$.} \label{sec:actualcalc}

First, we calculate the quantities $(x_i,y_i,z_i)$ for each of the springs shown in Fig. \ref{fig:vertices}. Then, we expand the energy to first and second order in the $(x_i,y_i,z_i)$. The resulting expression is then expanded once again, now to second order in both the strain tensor $\stackrel{\leftrightarrow}{\epsilon}$ and the rotation angle $\theta$. As both quantities are expected to be small, we restrict our consideration to terms that are at most quadratic in either one or both of them. The next step is to determine the angle $\psi$, (shown in Fig. \ref{fig:angles}) by which the lattice shears. This angle is again controlled by the requirement that the lattice, in the absence of any imposed stress, can be stabilized by an osmotic pressure $\Pi$, that couples only to the total volume of the lattice, or in other words to the combination $\mathop{\rm Tr} \stackrel{\leftrightarrow}{\epsilon} = \epsilon_{xx} + \epsilon_{yy} + \epsilon_{zz}$. 

If we expand the energy of the lattice to first order in $\stackrel{\leftrightarrow}{\epsilon}$ and zeroth order in $\theta$, we obtain the following expression
\begin{equation}
3a^2 ( \epsilon_{xx} + \epsilon_{yy}) \cos^2 \psi + 6 a \epsilon_{zz} \sin \psi ( a \sin \psi - 2l)
\label{eq:linear1}
\end{equation}
The requirement that this reduce to a function of the trace of the strain tensor only leads to a quadratic equation for $\sin \psi$. The solution to this equation is
\begin{eqnarray}
\psi & = &  \arcsin \left[ \frac{2l + \sqrt{3a^2+4l^2}}{3a}\right] \nonumber \\
& \equiv & \Psi_0(l/a)
\label{eq:linear2}
\end{eqnarray}
It is useful to consider the two limits of the above expression. When $l \rightarrow 0$, the argument of the arcsin reduces to $1/\sqrt{3}$, which is consistent with an arctangent equal to $1/\sqrt{2}$ and a cubic lattice. On the other hand, as $l \rightarrow a/2$, the angle $\psi$ goes to $\pi/2$, which means that the lattice collapses onto itself. In the latter limit, the rods are long enough that they touch end-to-end when that collapse takes place. 

Inserting the value of $\psi$ given by the right hand side of (\ref{eq:linear2}) into the expression (\ref{eq:linear1}), we find for the dependence of the energy on uniform dilations of the lattice
\begin{equation}
\frac{2}{3} \left( \epsilon_{xx} + \epsilon_{yy} + \epsilon_{zz} \right) \left[ 3a^2 - 4l^2 -2l \sqrt{3a^2+4l^2}\right]
\label{eq:linear3}
\end{equation}
so the Equation of State of the system is
\begin{equation}
\Pi v_0 = \frac{2}{3} \nu_1  \left[ 3a^2 - 4l^2 -2l \sqrt{3a^2+4l^2}\right]
\label{eq:eos}
\end{equation}
where 
\begin{eqnarray}
v_0 & = &  a^3 (3 \sqrt{3}/2) \cos ^2\psi  \sin \psi  \nonumber \\
& = & \frac{(3 a^2 -8l^2)\sqrt{3 a^2+4 l^2}-16 l^3 }{3 \sqrt{3} }
\label{eq:v0form}
\end{eqnarray}
is the volume per rod in the lattice. 
As $l \rightarrow a/2$, the expression multiplying the trace of the strain tensor vanishes, and the linear dependence of the energy on uniform dilations of the lattice is lost. The osmotic pressure, $\Pi$ approaches the limiting value $2\nu_1/\sqrt{3}a$. 

In light of this initial adjustment of the lattice, the next step is to determine the extent to which the lattice responds to an external stress. The response manifests itself in three quantities: the strain tensor $\stackrel{\leftrightarrow}{\epsilon}$, the rod tilt angle $\theta$ and the azimuthal angle, $\phi$, of the rods. As it turns out, the azimuthal angle appears only in the term that is first order in $\theta$, and the form of that term is $A \cos \phi + B \sin \phi$, where $A$ and $B$ are linear functions of the strain tensor and general functions of the angle $\psi$. This expression is minimized when $\phi = \arctan A/B$ and is equal in this case to $- \sqrt{A^2+B^2}$. The second-order-in-$\theta$ contribution to the energy of the deformed lattice is equal to
\begin{equation}
6al \left[ 2al\nu_2 \left( \cos \Psi_0 (l/a) \right)^2 + \nu_1 \sin \Psi_0(l/a) \right] \theta^2 \equiv D_2 \theta^2
\label{eq:theta2}
\end{equation}
If we denote by $D_0$ the portion of the deformation enerrgy that is zeroth order in $\theta$ and second order in the strain tensor, the net free energy as a function of the polar angle is
\begin{equation}
D_0 - \theta \sqrt{A^2+B^2} + D_2 \theta^2
\label{eq:thetas}
\end{equation}
Minimizing (\ref{eq:thetas}) with respect to $\theta$, we end up with our final result for the dependence of the energy on the strain tensor $\stackrel{\leftrightarrow}{\epsilon}$ in the uniformly strained lattice. In terms of the quantities defined above, the energy has the form 
\begin{eqnarray}
D_0 - \frac{A^2+B^2}{4D_2} & = & \sum_{i,j,k,l} \alpha_{ij,kl} \epsilon_{ij} \epsilon_{kl} \nonumber \\
& \equiv & \Omega \left[ \stackrel{\leftrightarrow}{\epsilon}\right]
\label{eq:enform}
\end{eqnarray}
As indicated in (\ref{eq:enform}), the energy of the uniformly distorted lattice with the angles $\psi$, $\phi$ and $\theta$ relaxed to the values dictated by energy minimization and lattice stabilization is purely quadratic in the strain tensor. The tensorial quantity $\alpha_{ij, kl}$ is symmetric with respect to interchange of the index pairs, $ij$ and $kl$. The response of the system of rods and springs to an externally generated stress tensor, $\stackrel{\leftrightarrow}{\sigma}$, follows from the solution to the set of linear equations
\begin{equation}
\frac{\partial \Omega \left[\stackrel{\leftrightarrow}{\epsilon} \right]}{\partial \epsilon_{ij}} = \sigma_{ij}
\label{eq:respeq}
\end{equation}
Given the overall anisotropy of the lattice, symmetry arguments yield only restricted information. If we orient the $z$ axis along the direction defined by the rods in the unstressed  lattice, then symmetry arguments tell us that a stress entirely in the $z$ direction results in equal values of the strain tensor components $\epsilon_{xx}$ and $\epsilon_{yy}$, while $\epsilon_{xy}=\epsilon_{yx} =0$. 

\subsection{Poisson ratios for $l=0$}

\begin{figure}[htbp]
\begin{center}
\includegraphics[width=3in]{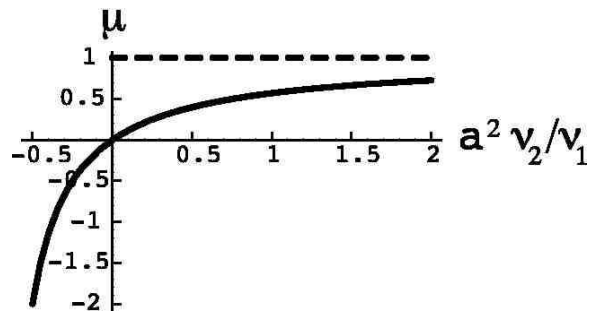}
\caption{The Poisson ratio as a function of $a^2 \nu_2/\nu_1$ when $l=0$. The asymptote of one is also indicated.}
\label{fig:l0poiss}
\end{center}
\end{figure}

Once again we consider the case of a lattice consisting of springs only, in which the rods are replaced by  point connections. For $l=0$, the Poisson ratio of interest becomes
\begin{equation}
\mu=\frac{4(a^2 \nu_2/\nu_1)}{3+4(a^2 \nu_2/\nu_1)}
\label{eq:l0poiss}
\end{equation}
and the determinant of the energy matrix is 
\begin{equation}
\det \stackrel{\leftrightarrow}{\alpha} = 512 a^{18}\nu_1^9\left(1 +\frac{a^2\nu_2}{\nu_1}\right)^3 
\label{eq:l0det}
\end{equation}
This means the energy matrix is associated with a stable equilibrium state as long as $a^2 \nu_2/\nu_1 >-1/2$. Figure \ref{fig:l0poiss}  is a plot of the Poisson ratio as given by (\ref{eq:l0poiss}) in the range of stability of the energy.
The limits of this quantity are 1 as $a^2 \nu_2/\nu_1 \rightarrow \infty$ and -2 as $a^2 \nu_2/\nu_1 \rightarrow -1/2$, which are the limits of the Poisson ratio in an isotropic solid. This means that, as in two dimensions, broken rotational symmetry is not essential for the negative Poisson ratio.

\subsection{Poisson ratios for $l \neq 0$} \label{sec:powdep}

\begin{figure}[htbp]
\begin{center}
\includegraphics[width=3in]{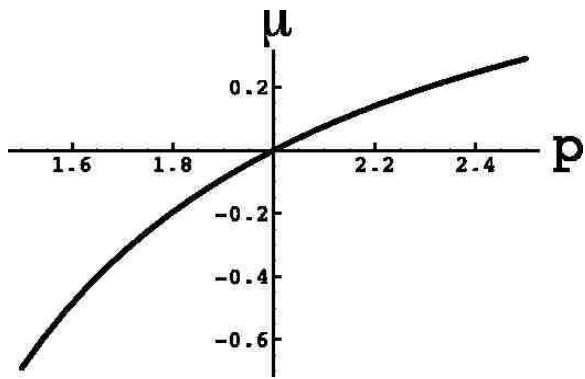}
\caption{Graph of the Poisson ratio versus the exponent, $p$, of the power law in (\ref{eq:interaction1}).}
\label{fig:poissgraph3}
\end{center}
\end{figure}

To explore the properties of the Poisson ratio for $l \neq 0$, we will focus on the exponent $p$ in the power law interaction  (\ref{eq:interaction1}). Consider the graph shown in Fig. \ref{fig:poissgraph3}, which is of the Poisson ratio, defined as
\begin{equation}
\mu = - \frac{\epsilon_{xx} + \epsilon_{yy}}{\epsilon_{zz}}
\label{eq:poiss1}
\end{equation}
in the case of a stress entirely in the $z$ direction ($\sigma_{ij}=0$ except for $\sigma_{zz}$). The length, $l$, of the rod has been set equal to $0.4a$. 
Note that the Poisson ratio passes through zero as the power law passes through the power associated with the Gaussian spring. When $p<2$, the Poisson ratio is negative, which means  that the lattice system resists shearing. 

\begin{figure}[htbp]
\begin{center}
\includegraphics[width=3in]{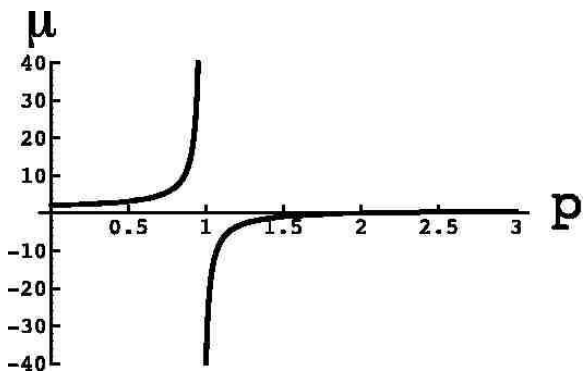}
\caption{The Poisson ratio, plotted against the exponent $p$ in (\ref{eq:interaction1}). Here $p$ ranges from zero to three. As noted in the text, a value of $p$ less than one is inconsistent with mechanical stability.}
\label{fig:poissvp}
\end{center}
\end{figure}

Figure \ref{fig:poissvp} is a graph of the Poisson ratio when $l=0.45 a$ as a function of $p$. 
The range of stability, i.e. of a positive determinant for $\stackrel{\leftrightarrow}{\alpha}$, is for $p>1$, with $p=1$ the location of the onset of instability. Recall that in the two dimensional case, $\mu$ approaches a finite negative value at the point of instability. From Fig. \ref{fig:poissvp} we see that the Poisson ratio \emph{diverges to minus infinity at the instability threshold}. The amplification effect of the rotational degree of freedom in producing a large negative Poisson ratio is thus much more pronounced in three dimensions. By ``tuning'' the physical properties of the non-Gaussian springs it is possible to dramatically alter the elastic response of the network.

\section{RNA in networks and a negative Poisson ratio}

\begin{figure}[htbp]
\begin{center}
\includegraphics[width=3in]{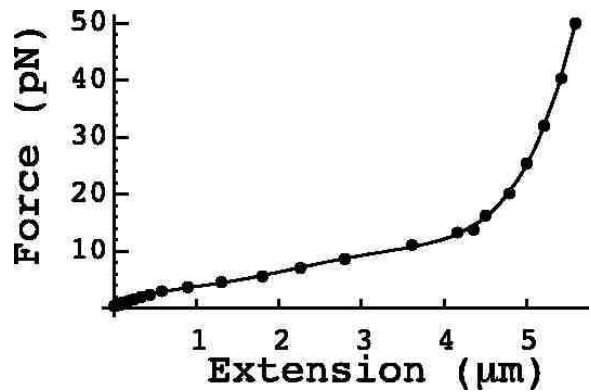}
\caption{Force versus extension data for a segment of single-stranded charomid DNA \cite{saito} with a backbone length of 5.7 $\mu$m in one millimolar phosphate buffer. The data were extracted from Figure 1 of \cite{dessinges}. Also shown in this figure is a fifth order polynomial fit to the data.}
\label{fig:data}
\end{center}
\end{figure}

\begin{figure}[htbp]
\begin{center}
\includegraphics[width=3in]{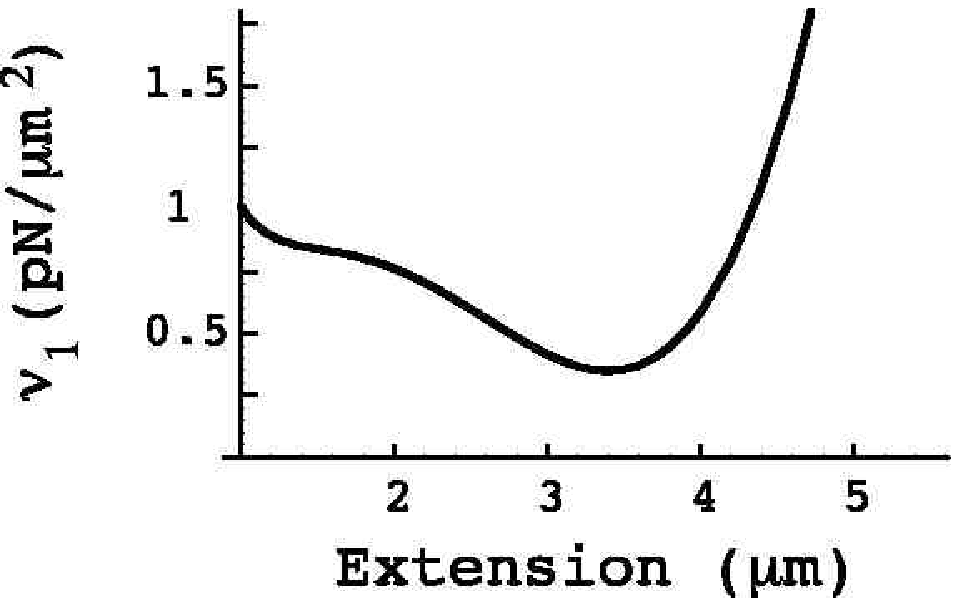}
\caption{The quantity $\nu_1$ as defined in (\ref{eq:expansion}) and as derived from the fitting curve in Fig. \ref{fig:data}. }
\label{fig:nu1plot}
\end{center}
\end{figure}

\begin{figure}[htbp]
\begin{center}
\includegraphics[width=3in]{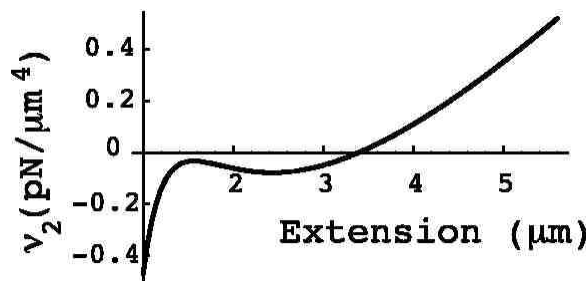}
\caption{The quantity $\nu_2$ as defined in (\ref{eq:expansion}) and as derived from the fitting curve in Fig. \ref{fig:data}. }
\label{fig:nu2plot}
\end{center}
\end{figure}

In this conclusion, we will apply the results of the last section to RNA networks and discuss the consequences in terms of soft elasticity and negative Poisson ratios.. In Sections \ref{sec:2d} and \ref{sec:3d} we learned that $\nu_2$ functions as a control parameter for elasticity. If $\nu_2$ is positive, one obtains conventional elasticity. As noted in the Introduction, gels of flexible and semi-flexible polymers are conventionally modeled as networks of Gaussian chains, worm-like chains or freely jointed chains, all of which have $\nu_2 >0$. If $\nu_2$ is negative, we predict a range of anomalous elasticity, with negative Poisson ratios terminating in a mechanical instability at a critical value for $\nu_2$, at which point $\mu$ diverges to minus infinity. Is it realistic for a biopolymer network to have a negative $\nu_2$? The $\nu_2$ parameter can in principle be determined from the force-extension curves of the nonlinear springs. For an RNA network of the type shown in Fig. \ref{fig:rodsandsprings}, this would be the force-extension curve of single stranded RNA. Force-extension curves have been measured for single-straned RNA and folded RNA molecules \cite{williams} but not yet for long single-stranded chains. Such measurements have been performed for single-stranded DNA strands of about $10^4$ bases \cite{dessinges}. A typical set of results, taken from \cite{dessinges} is shown in Fig. \ref{fig:data}. The force-extension curve rises rapidly when the extension reaches the contour length. This is preceded by a range of extensions for which the force is relatively constant, which may be due to progressive loss of stacking interaction. The force-extension curve for small extensions is presumably dominate by entropic elasticity. We fitted the measured data with a fifth order polynomial form, from which we were able to extract the quantities $\nu_1$ (in pN/$\mu$m$^2$) and $\nu_2$ (in pN/$\mu$m$^4$), shown in Figs. \ref{fig:nu1plot} and \ref{fig:nu2plot}.

\begin{figure}[htbp]
\begin{center}
\includegraphics[width=3in]{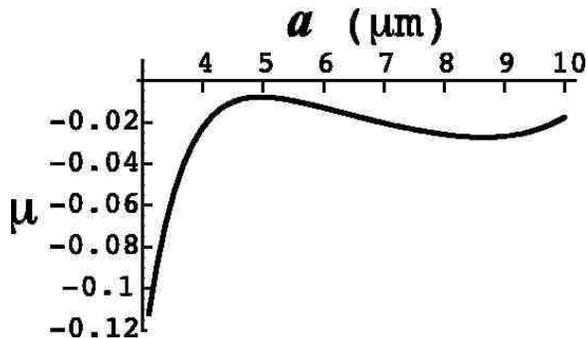}
\caption{The Poisson ratio, $\mu$,  derived from the parameters $\nu_1$ and $\nu_2$ displayed in Figs. \ref{fig:nu1plot} and \ref{fig:nu2plot}. The ratio $l/a$ is fixed at the value 0.45. Note that the parameter $a$  on the horizontal axis of this plot refers to the distance between rod centers and does not coincide with the extension shown in Figs. \ref{fig:nu1plot} and \ref{fig:nu2plot}. }
\label{fig:realpoiss}
\end{center}
\end{figure}

The $\nu_2$ control parameter is indeed negative for extensions less than 3 $\mu$m. It should be noted that $\nu_1$ and $\nu_2$ in general are expected to be quite sensitive to solvent conditions. We can now make use of the methods described above to compute the Poisson ratio for given ratio $l/a$, and we find that there is indeed a negative Poisson ratio over the full range of values of the length parameter $a$, the value of $l/a$ having been fixed at 0.45 (see Fig. \ref{fig:realpoiss}).

These results are, of course, merely illustrative, but they demonstrate that RNA should be a beautiful ``laboratory system'' for the study of gels with anomalous elastic properties. As Fig. \ref{fig:nu2plot} shows, one can tune the control parameter $\nu_2$ by adjusting the extension, which can, in turn, be achieved via changes in the osmotic swelling pressure of the system. The second key parameter, the $l/a$ ratio, can be ``programmed'' into the RNA molecules by alternating random sequences with complementary homopolymer sequences of prescribed length (e.g. strings of C monomers alternating with strings of G monomers). The most challenging feature would be to generate alignment between the rods to produce a nematic elastomer. Curiously, large ss RNA molecule n the form of 1,400 base-long viral genomes are found to be quite anisotropic according to low angle X-ray diffraction and light scattering studies \cite{ribitsch}. This suggests that large ss RNA molecules may be \emph{naturally} anisotropic. Assembly of the gel under mild shear flow may enhance this natural anisotropy. In summary, RNA gels are expected to be rich laboratory systems for the study of fundamental elasticity.

\subsection{Implications for viral assembly}

\begin{figure}[htbp]
\begin{center}
\includegraphics[width=3in]{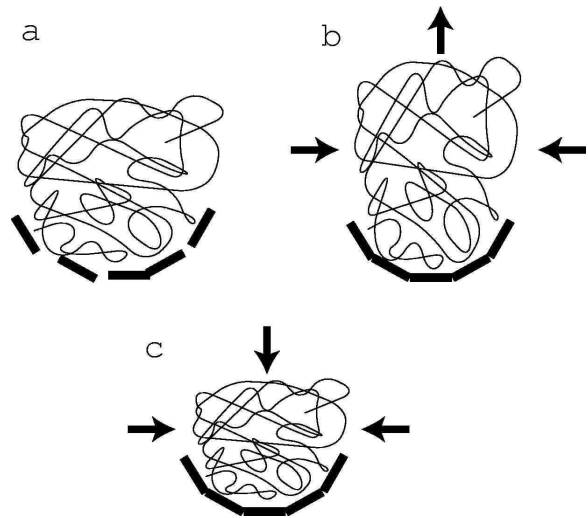}
\caption{The first steps in the assembly of a spherical RNA virus. In Fig. \ref{fig:compression}a, the capsomeres are beginning to come together. They attach to the portion of the genome in immediate proximity to them. Figure \ref{fig:compression}b shows what happens to the RNA complex as the capsomeres assemble and the RNA is compressed horizontally if the elastic response of the complex is characterized by a positive Poisson ratio. Figure \ref{fig:compression}c shows what happens if the RNA complex has a negative Poisson ratio.}
\label{fig:compression}
\end{center}
\end{figure}

It is interesting to speculate about possible consequences of a negative Poisson ratio for RNA gels. A curious problem of viral assembly is the large discrepancy between the density of the genome in solution and the same genome encapsidated by the virus's protein shell. Encapsidation proceeds by {\em self-assembly}: a solution of capsid proteins and viral RNA molecules will spontaneously assemble to form infectious viruses under physiological condittions. The assembly is driven by generic electrostatic affinity between the proteins and the RNA molecules, though specific interactions are required to initiate assembly see, for instance \cite{zlotnick}.  The scenario for the encapsidation of single-stranded RNA genomes into spherical viruses is not fully understood, but Fig. \ref{fig:compression} indicates
a likely scenario. A partially condensed RNA molecule forms a condensation surface for capsid proteins of opposite (i.e. positive) charge, and a curved shell starts to form. The challenge for the assembly process is how compaction of the genome can be achieved. If the RNA material had a positive Poisson ratio, then compaction of the genome at the nucleo-protein interface by the electrostatic affinity would produce \emph{swelling} of the remainder of the  molecule that is not in close contact with the growing shell. It would seem that completion of assembly is not possible. A negative Poisson ratio, on the other hand, would lead to collapse of the RNA material into the shell. It would be fascinating if model systems could be developed with, for instance, RNA gels in contact with a positively charged substrate to verify this scenario. We conclude that RNA gels with negative Poisson ratios would be well-adapted for easy encapsidation. 

\acknowledgements

The authors gratefully acknowledge support of the National Science Foundation through NSF grant DMR 0404507.

\bibliography{poisson}

\begin{thebibliography}{28}
\expandafter\ifx\csname natexlab\endcsname\relax\def\natexlab#1{#1}\fi
\expandafter\ifx\csname bibnamefont\endcsname\relax
  \def\bibnamefont#1{#1}\fi
\expandafter\ifx\csname bibfnamefont\endcsname\relax
  \def\bibfnamefont#1{#1}\fi
\expandafter\ifx\csname citenamefont\endcsname\relax
  \def\citenamefont#1{#1}\fi
\expandafter\ifx\csname url\endcsname\relax
  \def\url#1{\texttt{#1}}\fi
\expandafter\ifx\csname urlprefix\endcsname\relax\def\urlprefix{URL }\fi
\providecommand{\bibinfo}[2]{#2}
\providecommand{\eprint}[2][]{\url{#2}}

\bibitem[{\citenamefont{Flory}(1969)}]{flory}
\bibinfo{author}{\bibfnamefont{P.~J.} \bibnamefont{Flory}},
  \emph{\bibinfo{title}{Statistical Mechanics of Chain Molecules}}
  (\bibinfo{publisher}{Interscience}, \bibinfo{address}{New York},
  \bibinfo{year}{1969}).

\bibitem[{\citenamefont{Flory}(1947)}]{flory1}
\bibinfo{author}{\bibfnamefont{P.~J.} \bibnamefont{Flory}},
  \bibinfo{journal}{Journal Of Chemical Physics} \textbf{\bibinfo{volume}{15}},
  \bibinfo{pages}{397} (\bibinfo{year}{1947}).

\bibitem[{\citenamefont{Rubinstein and Colby}(2003)}]{rubinstein}
\bibinfo{author}{\bibfnamefont{M.}~\bibnamefont{Rubinstein}} \bibnamefont{and}
  \bibinfo{author}{\bibfnamefont{R.~H.} \bibnamefont{Colby}},
  \emph{\bibinfo{title}{Polymer Physics}} (\bibinfo{publisher}{Oxford
  University Press}, \bibinfo{address}{Oxford, New York},
  \bibinfo{year}{2003}).

\bibitem[{\citenamefont{Gardel et~al.}(2004)\citenamefont{Gardel, Shin,
  MacKintosh, Mahadevan, Matsudaira, and Weitz}}]{gardel}
\bibinfo{author}{\bibfnamefont{M.~L.} \bibnamefont{Gardel}},
  \bibinfo{author}{\bibfnamefont{J.~H.} \bibnamefont{Shin}},
  \bibinfo{author}{\bibfnamefont{F.~C.} \bibnamefont{MacKintosh}},
  \bibinfo{author}{\bibfnamefont{L.}~\bibnamefont{Mahadevan}},
  \bibinfo{author}{\bibfnamefont{P.}~\bibnamefont{Matsudaira}},
  \bibnamefont{and} \bibinfo{author}{\bibfnamefont{D.~A.} \bibnamefont{Weitz}},
  \bibinfo{journal}{Science} \textbf{\bibinfo{volume}{304}},
  \bibinfo{pages}{1301} (\bibinfo{year}{2004}).

\bibitem[{\citenamefont{Thirumalai and Hyeon}(2005)}]{thirumalai}
\bibinfo{author}{\bibfnamefont{D.}~\bibnamefont{Thirumalai}} \bibnamefont{and}
  \bibinfo{author}{\bibfnamefont{C.}~\bibnamefont{Hyeon}},
  \bibinfo{journal}{Biochemistry} \textbf{\bibinfo{volume}{44}},
  \bibinfo{pages}{4957} (\bibinfo{year}{2005}).

\bibitem[{\citenamefont{Isambert and Siggia}(2000)}]{isambert}
\bibinfo{author}{\bibfnamefont{H.}~\bibnamefont{Isambert}} \bibnamefont{and}
  \bibinfo{author}{\bibfnamefont{E.~D.} \bibnamefont{Siggia}},
  \bibinfo{journal}{Proceedings Of The National Academy Of Sciences Of The
  United States Of America} \textbf{\bibinfo{volume}{97}},
  \bibinfo{pages}{6515} (\bibinfo{year}{2000}).

\bibitem[{\citenamefont{Warner and Kutter}(2002)}]{warnerkutter}
\bibinfo{author}{\bibfnamefont{M.}~\bibnamefont{Warner}} \bibnamefont{and}
  \bibinfo{author}{\bibfnamefont{S.}~\bibnamefont{Kutter}},
  \bibinfo{journal}{Physical Review E} \textbf{\bibinfo{volume}{65}}
  (\bibinfo{year}{2002}), \bibinfo{note}{and references therein}.

\bibitem[{\citenamefont{Golubovic and Lubensky}(1989)}]{golublubensk}
\bibinfo{author}{\bibfnamefont{L.}~\bibnamefont{Golubovic}} \bibnamefont{and}
  \bibinfo{author}{\bibfnamefont{T.~C.} \bibnamefont{Lubensky}},
  \bibinfo{journal}{Physical Review Letters} \textbf{\bibinfo{volume}{63}},
  \bibinfo{pages}{1082} (\bibinfo{year}{1989}).

\bibitem[{hob()}]{hoberman}
\bibinfo{note}{See http://www.hoberman.com/fold/main/index.htm}.

\bibitem[{not()}]{note}
\bibinfo{note}{In the context of the elasticity of an isotropic solid, a
  negative Poisson ratio represents the dominance of the tendency of a body to
  retain its shape over the tendency to conserve its volume. See \cite{landl}.}

\bibitem[{\citenamefont{Love}(1944)}]{love}
\bibinfo{author}{\bibfnamefont{A.~E.~H.} \bibnamefont{Love}},
  \emph{\bibinfo{title}{A Treatise on the Mathematical Theory of Elasticity}}
  (\bibinfo{publisher}{Dover}, \bibinfo{address}{New York},
  \bibinfo{year}{1944}).

\bibitem[{\citenamefont{Bowick et~al.}(2001)\citenamefont{Bowick, Cacciuto,
  Thorleifsson, and Travesset}}]{bowick}
\bibinfo{author}{\bibfnamefont{M.}~\bibnamefont{Bowick}},
  \bibinfo{author}{\bibfnamefont{A.}~\bibnamefont{Cacciuto}},
  \bibinfo{author}{\bibfnamefont{G.}~\bibnamefont{Thorleifsson}},
  \bibnamefont{and}
  \bibinfo{author}{\bibfnamefont{A.}~\bibnamefont{Travesset}},
  \bibinfo{journal}{Physical Review Letters} \textbf{\bibinfo{volume}{87}},
  \bibinfo{pages}{148103/1} (\bibinfo{year}{2001}).

\bibitem[{\citenamefont{Lubarda and Meyers}(1999)}]{lubarda}
\bibinfo{author}{\bibfnamefont{V.~A.} \bibnamefont{Lubarda}} \bibnamefont{and}
  \bibinfo{author}{\bibfnamefont{M.~A.} \bibnamefont{Meyers}},
  \bibinfo{journal}{Scripta Materialia} \textbf{\bibinfo{volume}{40}},
  \bibinfo{pages}{975} (\bibinfo{year}{1999}).

\bibitem[{\citenamefont{Yuejin and Goddard}(1995)}]{yuejin}
\bibinfo{author}{\bibfnamefont{G.}~\bibnamefont{Yuejin}} \bibnamefont{and}
  \bibinfo{author}{\bibfnamefont{I.}~\bibnamefont{Goddard},
  \bibfnamefont{W.~A.}}, \bibinfo{journal}{Chemical Physics Letters}
  \textbf{\bibinfo{volume}{237}}, \bibinfo{pages}{72} (\bibinfo{year}{1995}).

\bibitem[{\citenamefont{Brandel and Lakes}(2001)}]{brandel}
\bibinfo{author}{\bibfnamefont{B.}~\bibnamefont{Brandel}} \bibnamefont{and}
  \bibinfo{author}{\bibfnamefont{R.~S.} \bibnamefont{Lakes}},
  \bibinfo{journal}{Journal of Materials Science}
  \textbf{\bibinfo{volume}{36}}, \bibinfo{pages}{5885} (\bibinfo{year}{2001}).

\bibitem[{\citenamefont{Gaspar et~al.}(2005)\citenamefont{Gaspar, Ren, Smith,
  Grima, and Evans}}]{gaspar}
\bibinfo{author}{\bibfnamefont{N.}~\bibnamefont{Gaspar}},
  \bibinfo{author}{\bibfnamefont{X.~J.} \bibnamefont{Ren}},
  \bibinfo{author}{\bibfnamefont{C.~W.} \bibnamefont{Smith}},
  \bibinfo{author}{\bibfnamefont{J.~N.} \bibnamefont{Grima}}, \bibnamefont{and}
  \bibinfo{author}{\bibfnamefont{K.~E.} \bibnamefont{Evans}},
  \bibinfo{journal}{Acta Materialia} \textbf{\bibinfo{volume}{53}},
  \bibinfo{pages}{2439} (\bibinfo{year}{2005}).

\bibitem[{\citenamefont{Grima et~al.}(2005)\citenamefont{Grima, Alderson, and
  Evans}}]{grima}
\bibinfo{author}{\bibfnamefont{J.~N.} \bibnamefont{Grima}},
  \bibinfo{author}{\bibfnamefont{A.}~\bibnamefont{Alderson}}, \bibnamefont{and}
  \bibinfo{author}{\bibfnamefont{K.~E.} \bibnamefont{Evans}},
  \bibinfo{journal}{Physica Status Solidi B-Basic Solid State Physics}
  \textbf{\bibinfo{volume}{242}}, \bibinfo{pages}{561} (\bibinfo{year}{2005}).

\bibitem[{\citenamefont{Evans et~al.}(1992)\citenamefont{Evans, Nkansah, and
  Hutchinson}}]{evans}
\bibinfo{author}{\bibfnamefont{K.~E.} \bibnamefont{Evans}},
  \bibinfo{author}{\bibfnamefont{M.~A.} \bibnamefont{Nkansah}},
  \bibnamefont{and} \bibinfo{author}{\bibfnamefont{I.~J.}
  \bibnamefont{Hutchinson}}, \bibinfo{journal}{Acta Metallurgica et Materialia}
  \textbf{\bibinfo{volume}{40}}, \bibinfo{pages}{2463} (\bibinfo{year}{1992}).

\bibitem[{\citenamefont{Veronda and Westmann}(1970)}]{skin}
\bibinfo{author}{\bibfnamefont{D.~R.} \bibnamefont{Veronda}} \bibnamefont{and}
  \bibinfo{author}{\bibfnamefont{R.~A.} \bibnamefont{Westmann}},
  \bibinfo{journal}{Journal Of Biomechanics} \textbf{\bibinfo{volume}{3}},
  \bibinfo{pages}{111} (\bibinfo{year}{1970}).

\bibitem[{\citenamefont{Williams and Lewis}(1982)}]{bone}
\bibinfo{author}{\bibfnamefont{J.~L.} \bibnamefont{Williams}} \bibnamefont{and}
  \bibinfo{author}{\bibfnamefont{J.~L.} \bibnamefont{Lewis}},
  \bibinfo{journal}{Journal Of Biomechanical Engineering-Transactions Of The
  Asme} \textbf{\bibinfo{volume}{104}}, \bibinfo{pages}{50}
  (\bibinfo{year}{1982}).

\bibitem[{ins()}]{instabnote}
\bibinfo{note}{The instability is associated with an anisotropic expansion or
  contraction of the lattice in which $\epsilon_{xy}= \epsilon_{yx}=0$ and
  $\epsilon_{xx}/\epsilon_{yy}= 8.42452$.}

\bibitem[{\citenamefont{Ting and Chen}(2005)}]{ting}
\bibinfo{author}{\bibfnamefont{T.~C.~T.} \bibnamefont{Ting}} \bibnamefont{and}
  \bibinfo{author}{\bibfnamefont{T.~Y.} \bibnamefont{Chen}},
  \bibinfo{journal}{Quarterly Journal Of Mechanics And Applied Mathematics}
  \textbf{\bibinfo{volume}{58}}, \bibinfo{pages}{73} (\bibinfo{year}{2005}),
  \bibinfo{note}{part 1}.

\bibitem[{\citenamefont{Williams and Rouzina}(2002)}]{williams}
\bibinfo{author}{\bibfnamefont{M.~C.} \bibnamefont{Williams}} \bibnamefont{and}
  \bibinfo{author}{\bibfnamefont{I.}~\bibnamefont{Rouzina}},
  \bibinfo{journal}{Current Opinion In Structural Biology}
  \textbf{\bibinfo{volume}{12}}, \bibinfo{pages}{330} (\bibinfo{year}{2002}).

\bibitem[{\citenamefont{Dessinges et~al.}(2002)\citenamefont{Dessinges, Maier,
  Zhang, Peliti, Bensimon, and Croquette}}]{dessinges}
\bibinfo{author}{\bibfnamefont{M.~N.} \bibnamefont{Dessinges}},
  \bibinfo{author}{\bibfnamefont{B.}~\bibnamefont{Maier}},
  \bibinfo{author}{\bibfnamefont{Y.}~\bibnamefont{Zhang}},
  \bibinfo{author}{\bibfnamefont{M.}~\bibnamefont{Peliti}},
  \bibinfo{author}{\bibfnamefont{D.}~\bibnamefont{Bensimon}}, \bibnamefont{and}
  \bibinfo{author}{\bibfnamefont{V.}~\bibnamefont{Croquette}},
  \bibinfo{journal}{Physical Review Letters} \textbf{\bibinfo{volume}{89}}
  (\bibinfo{year}{2002}).

\bibitem[{\citenamefont{Saito and Stark}(1986)}]{saito}
\bibinfo{author}{\bibfnamefont{I.}~\bibnamefont{Saito}} \bibnamefont{and}
  \bibinfo{author}{\bibfnamefont{G.~R.} \bibnamefont{Stark}},
  \bibinfo{journal}{Proceedings Of The National Academy Of Sciences Of The
  United States Of America} \textbf{\bibinfo{volume}{83}},
  \bibinfo{pages}{8664} (\bibinfo{year}{1986}).

\bibitem[{\citenamefont{Ribitsch et~al.}(1985)\citenamefont{Ribitsch, Declercq,
  Folkhard, Zipper, Schurz, and Clauwaert}}]{ribitsch}
\bibinfo{author}{\bibfnamefont{G.}~\bibnamefont{Ribitsch}},
  \bibinfo{author}{\bibfnamefont{R.}~\bibnamefont{Declercq}},
  \bibinfo{author}{\bibfnamefont{W.}~\bibnamefont{Folkhard}},
  \bibinfo{author}{\bibfnamefont{P.}~\bibnamefont{Zipper}},
  \bibinfo{author}{\bibfnamefont{J.}~\bibnamefont{Schurz}}, \bibnamefont{and}
  \bibinfo{author}{\bibfnamefont{J.}~\bibnamefont{Clauwaert}},
  \bibinfo{journal}{Zeitschrift Fur Naturforschung C-A Journal Of Biosciences}
  \textbf{\bibinfo{volume}{40}}, \bibinfo{pages}{234} (\bibinfo{year}{1985}).

\bibitem[{\citenamefont{Zlotnick et~al.}(2000)\citenamefont{Zlotnick, Aldrich,
  Johnson, Ceres, and Young}}]{zlotnick}
\bibinfo{author}{\bibfnamefont{A.}~\bibnamefont{Zlotnick}},
  \bibinfo{author}{\bibfnamefont{R.}~\bibnamefont{Aldrich}},
  \bibinfo{author}{\bibfnamefont{J.~M.} \bibnamefont{Johnson}},
  \bibinfo{author}{\bibfnamefont{P.}~\bibnamefont{Ceres}}, \bibnamefont{and}
  \bibinfo{author}{\bibfnamefont{M.~J.} \bibnamefont{Young}},
  \bibinfo{journal}{Virology} \textbf{\bibinfo{volume}{277}},
  \bibinfo{pages}{450} (\bibinfo{year}{2000}).

\bibitem[{\citenamefont{Landau et~al.}(1995)\citenamefont{Landau, Lifshitz,
  Kosevich, and Pitaevskii}}]{landl}
\bibinfo{author}{\bibfnamefont{L.~D.} \bibnamefont{Landau}},
  \bibinfo{author}{\bibfnamefont{E.~M.} \bibnamefont{Lifshitz}},
  \bibinfo{author}{\bibfnamefont{A.~M.} \bibnamefont{Kosevich}},
  \bibnamefont{and} \bibinfo{author}{\bibfnamefont{L.~P.}
  \bibnamefont{Pitaevskii}}, \emph{\bibinfo{title}{Theory of elasticity}}
  (\bibinfo{publisher}{Butterworth-Heinemann}, \bibinfo{address}{Oxford;
  Boston}, \bibinfo{year}{1995}), \bibinfo{edition}{3rd} ed.

\end{thebibliography}

 \end{document}